\documentclass[aps,prd,amsmath,amssymb,onecolumn]{revtex4}

\usepackage{graphicx}
\usepackage{dcolumn}
\usepackage{bm}
\usepackage{amsmath}
\usepackage{epstopdf}
\usepackage{amsfonts}
\usepackage{amssymb}
\usepackage{epsfig}
\usepackage{tabularx}
\usepackage{color}
\usepackage{soul}
\usepackage{multirow}
\usepackage{xcolor}

\usepackage[colorlinks=true,citecolor=blue,urlcolor=magenta,breaklinks]{hyperref}
\newcommand{\be}{\begin{equation}}
\newcommand{\en}{\end{equation}}
\newcommand{\bea}{\begin{eqnarray}}
\newcommand{\ena}{\end{eqnarray}}

\newcommand{\bff}{\begin{figure}}
\newcommand{\eff}{\end{figure}}

\begin{document}

\title{Universal relations for non-rotating objects made of dark energy}

\author{Grigoris Panotopoulos}

\address{
Departamento de Ciencias F{\'i}sicas, Universidad de la Frontera, Casilla 54-D, 4811186 Temuco, Chile.
\\
\href{mailto:grigorios.panotopoulos@ufrontera.cl}{\nolinkurl{grigorios.panotopoulos@ufrontera.cl}} 
}

\begin{abstract}
We obtain universal relations for fluid spheres without rotation made of dark energy assuming the Extended-Chaplygin Gas equation-of-state. After integrating the relevant differential equations, we make a fit to obtain the unknown coefficients of the functions a) normalized moment of inertia versus dimensionless deformability and b) normalized moment of inertia versus factor of compactness. We find that the form of the functions does not depend on the details of the underlying equation-of-state.
\end{abstract}

\maketitle

%%%%%%%%%%%%%%%%%%%%%%%
\section{Introduction}
%%%%%%%%%%%%%%%%%%%%%%%

During the circle of life of stars, a given star goes through different stages of evolution, starting from the main sequence until it eventually becomes a compact object. Compact stars \cite{Shapiro:1983du,Schaffner-Bielich:2020psc}, such as white dwarfs, neutron stars and black holes, comprise the final stages of stars, and they are characterized by very high matter densities. Neutron stars (NSs) in particular, with masses of approximately two solar masses and radii of around (10-15) kilometers, are fascinating objects since understanding their properties and observed complex phenomena (such as strong magnetic fields, hyperon-dominated matter etc) is a multidisciplinary task. It requires bringing together different scientific areas and lines of research, such as Nuclear Physics, Astrophysics and gravitational physics. Soon after the discovery of the neutron by James Chadwick \cite{Chadwick:1932ma, Chadwick:1932wcf}, Baade and Zwicky predicted that neutron stars should exist \cite{Baade:1934wuu}. One year after the discovery of pulsars in 1967 \cite{Hewish:1968bj}, their identification as NSs was established after the discoveries of pulsars in the Crab and Vela supernova remnants \cite{Safi-Harb:2017xso}. Those ultra-dense objects, thanks to their extreme conditions that cannot be reached on earth-based experiments, constitute excellent cosmic laboratories to study and constrain strongly interacting matter properties at high densities, phase transitions inside, non-conventional physics and alternative/modified theories of gravity. For a review on the physics of neutron stars see e.g. \cite{Lattimer:2004pg}.

\smallskip

It is speculated that less conventional, or even exotic, astronomical objects may exist. A new class of compact stars that may be an alternative to neutron stars is the so called strange quark stars (QSs) \cite{Haensel:1986qb, Itoh:1970uw, Collins:1974ky, Madsen1}. As their name suggests, they are supposed to be made of quark matter, and as of today they still remain hypothetical objects. According to the authors of \cite{Witten:1984rs,Farhi:1984qu}, quark matter could be the true ground state of hadrons, as by assumption is absolutely stable \cite{Witten:1984rs,Farhi:1984qu}. That property makes them a plausible explanation of some puzzling super-luminous supernovae \cite{Ofek:2006vt,Ouyed:2009dr}, which are much less frequent and at the same time brighter than regular supernovae. Indeed, they occur in about one out of every 1000 supernovae explosions, and they are more than 100 times more luminous than regular supernovae. In spite of the fact that as of today they remain hypothetical astronomical objects, strange QSs cannot conclusively be ruled out yet. In fact, there are some claims in the literature that there are currently some observed compact objects exhibiting peculiar features (e.g. small radii) that cannot be explained by the usual hadronic EoSs adopted in studies of NSs, see e.g. \cite{Henderson:2007gu, Li:2011zzn, Aziz:2019rgf}, and also Table 5 of \cite{Weber:2004kj} and references therein. 

\smallskip

It is currently estimated that the dominant component that drives the expansion of the Universe is dark energy (DE), corresponding to around 70$\%$ of its energy budget \cite{Planck:2018vyg}, responsible for the accelerating expansion of the Universe \cite{SupernovaSearchTeam:1998fmf, SupernovaCosmologyProject:1998vns,Freedman:2003ys}. Given that current cosmic acceleration calls for dark energy, in some recent works the authors proposed to study dark energy stars \cite{DEstars1,DEstars2,DEstars3,DEstars4,DEstars5} assuming an extended equation-of-state (EoS) of the form \cite{Pourhassan:2014cea, Pourhassan:2014ika}
\begin{equation}
p = - \frac{B^2}{\rho^a} + A^2 \rho,
\end{equation}
with $A, B$ and $a$ being constant parameters, where $p,\rho$ are the pressure and the density of the fluid, respectively. A simplified version of this, namely 
\begin{equation}
p = - \frac{B^2}{\rho^a},
\end{equation}
with $0 < a \leq 1$, known as generalized Chaplygin gas equation-of-state, was introduced in Cosmology long time ago in order to unify the description of non-relativistic matter together with a positive cosmological constant \cite{chaplygin1,chaplygin2}.

\smallskip

The inspiral and subsequent relativistic collisions of two stars in binaries, and the gravitational wave signal emitted during the whole process, contain a wealth of information regarding the nature of the colliding bodies. The imprint of the EoS within the signals emitted during binary coalescences is mainly determined by adiabatic tidal interactions, characterized by a set of coefficients, known as the tidal Love numbers and the corresponding deformabilities. The theory of tidal deformability was first introduced in Newtonian gravity by Love \cite{Love1,Love2} more than a century ago, with the purpose of understanding the yielding of the Earth to disturbing forces. In the case of a spherical body, Love introduced two dimensionless numbers to describe the tidal response of the Earth. To be more precise, the first number, $h$, describes the relative deformation of the body in the longitudinal direction (with respect to the perturbation), while the other one, $k$, describes the relative deformation of the gravitational potential. The consideration of self-gravitating compact objects requires a relativistic theory of tidal deformability, which was developed in \cite{flanagan,hinderer,damour,Lattimer,poisson} for spherically symmetric NSs/QSs and black holes. Naturally, the key deformability parameter is the relativistic generalization of $k$, since the role of the gravitational potential is now played by the metric tensor.

\smallskip

Stellar properties, such as mass and radius, strongly depend on the underlying EoS of matter content. As far as neutron stars is concerned, thanks to their very high matter densities and strong gravitational fields, NSs are considered to be excellent cosmic laboratories to constrain the poorly known hadronic equation-of-state as well as deviations from Einstein's General Relativity. As a matter of fact, it has been shown in the literature that M-R relationships obtained by different theories of gravity show a much greater variance than M-R relations obtained assuming different EoSs \cite{psaltis,tourkoi}. This implies that M-R measurements would constrain gravity models more than EoS models. Other methods to constrain the hadronic EoS and properties of dense matter are based on the moment of inertia \cite{inertia1,inertia2}, quasi-periodic oscillations \cite{andrea} and with electromagnetic and gravitational wave observations \cite{paper1,paper2}.

\smallskip

In an attempt to differenciate the inner structure and composition of NSs and QSs with respect to the EoS of dense matter and vice versa, certain EoS insensitive relations between stellar properties, such as mass, radius, moment of inertia, tidal deformability, quadrupole moment, fundamental mode of radial oscillations etc., have been found over the years. Those universal relations are both important and useful, as they would enable astronomers to infer some stellar properties from others that can be easily measured, even in the absence of precise information on the underlying EoS. For example, in \cite{inertia1,inertia2} the authors found formulas that relate the moment of inertia of strange stars and neutron stars to the stellar mass and radius. Moreover, the so called I-Love-Q universal relations, discovered in \cite{Yagi:2013bca,Yagi:2013awa}, relate the moment of inertia, $I$, to the spin-induced quadrupole moment, $Q$, and to the quadrupole tidal deformability, $\lambda_2$, expressing the distortion of a neutron star induced by tidal forces or spin in terms of its static structural parameters. For a partial list on universal relations in several contexts, such as alternative theories of gravity, rapidly rotating compact stars, including exotic matter and the mass etc., see e.g. \cite{UR1,UR2,UR3,UR4,UR5,UR6,UR7,UR8,UR9,UR10,UR11,UR12,UR13,UR14,UR15,UR16,UR17} and references therein.

\smallskip

Within General Relativity universal relations for more standard compact objects, such as neutron stars and strange quark stars, have been obtained. Therefore, here we shall be interested in less conventional spherical configurations. In the case of stars made of Bose-Einstein Condensate the EoS depends on the shape of the scalar self-interaction potential, although in the low energy density limit the EoS becomes approximately a polytrope with index 1, and this has been studied in \cite{Yagi:2013awa}.  For that reason in the present article we propose to obtain for the first time universal relations between (normalized) moment of inertia and tidal deformability for dark energy stars. To the best of our knowledge this has not been done so far, and therefore we wish to fill a gap in the literature. The plan of our work is the following: After this introductory section, in section II we briefly review the structure equations and tidal Love numbers of relativistic stars within Einstein's gravity. In the third section we present our main numerical results, and we finish in section IV discussing our findings and summarizing our work. Throughout the manuscript we adopt the mostly positive metric signature, and we work in geometrical units where Newton's constant and speed of light in vacuum are set to unity, $G=1=c$.

%%%%%%%%%%%%%%%%%%%%%%%%%%%%%%%%%%%%%%%%%%%%%%%%%%%%%%%%%%%%%%%%%
\section{Formalism: Structure equations and tidal deformability}
%%%%%%%%%%%%%%%%%%%%%%%%%%%%%%%%%%%%%%%%%%%%%%%%%%%%%%%%%%%%%%%%%

\subsection{Hydrostatic equilibrium}

Here we briefly review the structure equations for interior stellar solutions, starting from the field equations of Einstein's General Relativity \cite{Einstein:1915ca}
\begin{equation}
    G_{mn} \equiv R_{mn} - \frac{1}{2} \: R \: g_{mn} = 8 \pi  T_{mn},
\end{equation}
where $T_{mn}$ is the energy-momentum tensor of the matter content, $g_{mn}$ is the metric tensor, $R_{mn}$ and $R$ are the Ricci tensor and Ricci scalar, respectively, while $G_{mn}$ is the Einstein tensor.

The most general form of a static, spherically symmetric geometry in Schwarzschild-like coordinates, $\{ t, r, \theta, \phi \}$, is given by
\begin{equation}
    d s^2 = - e^{\nu} d t^2 + e^{\lambda} d r^2 + r^2 (d \theta^2 + \sin^2 \theta d \phi^2),
\end{equation}
where for interior solutions ($0 \leq r \leq R$, $R$ being the radius of the star) $\nu(r), \lambda(r)$ are two independent functions of the radial coordinate. In the discussion to follow it is more convenient to work with the mass function, $m(r)$, defined by
\begin{equation}
    \displaystyle e^{\lambda} = \frac{1}{1 - \frac{2 m}{r}}.
\end{equation}

To obtain interior solutions describing hydrostatic equilibrium of relativistic stars, one needs to integrate the Tolman-Oppenheimer-Volkoff (TOV) equations \cite{tolman, OV}
\begin{eqnarray}
    m'(r) & = & 4 \pi r^2 \rho (r),  \\ 
    p'(r) & = & - [ \rho(r) + p(r) ] \; \frac{\nu' (r)}{2}, \\
    \nu' (r) & = & 2 \: \frac{m(r) + 4 \pi r^3 p(r)}{ r^2 \left( 1 - 2 m(r) / r \right) },
\end{eqnarray}
where a prime denotes differentiation with respect to $r$. The isotropic matter content viewed as a perfect fluid is described by a stress-energy tensor of the form
\begin{equation}
T_a^b = Diag(-\rho, p, p, p),
\end{equation}
where $p$ is the pressure of the fluid, and $\rho$ is the energy density of matter content. Depending on the matter content, $p$ and $\rho$ satisfy a certain EoS $p(\rho)$.

To compute the stellar mass and radius, the two TOV equations
\begin{eqnarray}
    m'(r) & = & 4 \pi r^2 \rho (r),  \\ 
    p'(r) & = & - [ \rho(r) + p(r) ] \:  \frac{m(r) + 4 \pi r^3 p(r)}{ r^2 \left( 1 - 2 m(r) / r \right) } ,
\end{eqnarray}
together with the EoS are to be integrated imposing at the center of the star, $r \rightarrow 0$, appropriate initial conditions 
\begin{equation}
    m(0) = 0,
\end{equation}
\begin{equation}
    \rho(0) = \rho_{c},
\end{equation}
where $\rho_{c}$ is the central energy density. In addition, the following matching conditions must be satisfied at the surface of the object, $r \rightarrow R$
\begin{equation}
    p(R) = 0,
\end{equation}
\begin{equation}
    m(R) = M,
\end{equation}
\begin{equation}
    e^{\nu(R)} = 1-2 M/R,
\end{equation}
with $M$ being the stellar mass, taking into account that the exterior vacuum solution ($r > R$) is given by the Schwarzschild geometry \cite{Schwarzschild:1916uq}
\begin{equation}
    d s^2 = - (1-2M/r) d t^2 + (1-2M/r)^{-1} d r^2 + r^2 (d \theta^2 + \sin^2 \theta d \: \phi^2) .
\end{equation}

Finally, the first metric potential, $\nu(r)$, may be computed using the last TOV equation
\begin{equation}
\nu'(r)  =  2 \: \frac{m(r) + 4 \pi r^3 p(r)}{ r^2 \left( 1 - 2 m(r) / r \right) },
\end{equation}
supplemented by the third matching condition
\begin{equation}
    e^{\nu(R)} = 1-2 M/R,
\end{equation}
and therefore the solution is given by
\begin{equation}
    \displaystyle \nu (r) = \ln \left( 1 - \frac{2 M}{R} \right) + 2 \int^r_R \frac{m(x) + 4 \pi x^3 p_r(x)}{ x^2 \left( 1 - 2 m(x) / x \right) } \: dx .
\end{equation}

%%%%%%%%%

\subsection{Gravito-electric tidal Love numbers}

A complete description of the theory of tidal Love numbers can be consulted, e.g., in \cite{flanagan,hinderer,damour,Lattimer,poisson}. Let us consider a certain star subjected to an external gravitational field, $\Phi_{ext}$, produced, for instance, by a companion star in a binary.
The star under discussion will react to the external field by deforming. The leading deformation is to develop a quadrupolar moment $Q_{ij}$
\begin{equation}
Q_{ij} = \int d^3x \delta \rho(\vec{x}) \: (x_i x_j - \frac{1}{3} r^2 \delta_{ij}),
\end{equation}
which is proportional to the static external quadrupolar tidal field $E_{ij}$
\begin{equation}
Q_{ij} = - \lambda \: E_{ij},
\end{equation}
\begin{equation}
E_{ij} = \frac{\partial^2 \Phi_{ext}}{\partial x^i \partial x^j},
\end{equation}
and the spatial indices take three values $i,j=1,2,3$.

The tidal Love number $k$, a quadrupole moment number and dimensionless coefficient, depends on the structure of the star and thus on its mass and equation of state. It is directly related to two auxiliary quantities commonly referred to as "deformabilities", denoted $\lambda$ (dimensionful) and $\Lambda$ (dimensionless), which are defined as follows:
\begin{align}
\lambda &\equiv \frac{2}{3} k R^5,
\label{eq:Love1}
\\
\Lambda &\equiv \frac{2 k}{3 C^5},
\label{eq:Love2}
\end{align}
where $C=M/R$ is the well-known compactness factor of the star. 
Subsequently, the tidal Love number can be written in terms of $C$ as follows given by \cite{flanagan,hinderer,damour,Lattimer,poisson}
\begin{align}
k &= \frac{8C^5}{5} \: \frac{K_{o}}{3  \:K_{o} \: \ln(1-2C) + P_5(C)} ,
\label{elove}
\\
K_{o} &= (1-2C)^2 \: [2 C (y_R-1)-y_R+2] ,
\\
y_R &\equiv y(r=R) ,
\end{align}
where $P_5(C)$ is a fifth-order polynomial given by
\begin{align}
\begin{split}
P_5(C) = \: & 2 C \: \Bigl( 4 C^4 (y_R+1) + 2 C^3 (3 y_R-2) \ + 
\\
&
2 C^2 (13-11 y_R) + 3 C (5 y_R-8) -
\\
&
3 y_R + 6 \Bigl)  ,
\end{split}
\end{align}
%
% \begin{strip}
% \begin{equation}
% P_5(C) =  2 C \: \Bigl( 4 C^4 (y_R+1) + 2 C^3 (3 y_R-2) \ + 
% 2 C^2 (13-11 y_R) + 3 C (5 y_R-8) -3 y_R + 6 \Bigl) 
% \end{equation}
% \end{strip}
%
and the function $y(r)$ is the solution of a Riccati differential equation \cite{Lattimer}:
\begin{align}
\begin{split}
r y'(r) + y(r)^2 &+ y(r) e^{\lambda (r)} \Bigl[1 + 4 \pi r^2 ( p(r) - \rho (r) ) \Bigl] 
\\
&+ r^2 Q(r) = 0 ,
\end{split}
\end{align}
supplemented by the initial condition at the center, $r \rightarrow 0$, $y(0)=2$, where the function $Q(r)$, not to be confused with the tensor $Q_{ij}$, is given by
%**1
\begin{align}
\begin{split}
  \displaystyle Q(r) = 4 \pi e^{\lambda (r)} \Bigg[ 
  5 \rho (r) 
  &+ 9 p(r) + \frac{\rho (r) + p(r)}{c^2_s(r)} 
  \Bigg] 
  \\
  &- 6 \frac{e^{\lambda (r)}}{r^2} - \Bigl[\nu' (r)\Bigl]^2 .
\end{split}
\end{align}
Notice that $c_s^2 \equiv dp/d\rho = p'(r)/\rho'(r)$ is the speed of sound. 
%see e.g \cite{Lattimer} for more details.

It has been pointed out in \cite{damour,Lattimer,Hinderer:2009ca} that when phase transitions or density discontinuities are present, it is necessary to slightly modify the expressions above. Since in the present work the energy density takes a non-vanishing surface value, in our analysis we incorporated the following correction
\begin{equation}
y_R \rightarrow y_R - 3 \frac{\Delta \rho}{\langle \rho \rangle},
\end{equation}
where $\Delta \rho$ is the density discontinuity, and 
\begin{equation}
\langle \rho \rangle = \frac{3M}{4 \pi R^3},
\end{equation}
is the mean energy density throughout the object. 

\smallskip

It is easy to see that since $k \propto (1-2C)^2$, tidal Love numbers of black holes vanish due to the fact that the factor of compactness of black holes $C=1/2$. This is an intriguing result of classical GR saying that tidal Love numbers of black holes, as opposed to other types of compact objects, are precisely zero. Therefore, a measurement of a non-vanishing $k$ will be a smoking-gun deviation from the standard black hole of GR. Moreover, regarding gravity waves observatories, the Einstein Telescope will pin down very precisely the EoS of neutron stars \cite{Iacovelli:2023nbv}, while the Laser Interferometer Space Antenna is able to probe even extremely compact objects (with a factor of compactness $C > 1/3$), and it will set constraints on the Love numbers of highly-spinning central objects at $\sim 0.001-0.01$ level \cite{Piovano:2022ojl}.

%%%%%%%%%%%%%%%%%%%%%%%%%%%%%%%%%%%%%%%%%%%%%%
\section{Numerical analysis and main results}
%%%%%%%%%%%%%%%%%%%%%%%%%%%%%%%%%%%%%%%%%%%%%

In the discussion to follow, and in order to integrate the TOV equations, a certain equation-of-state must be assumed. Here we adopt the Extended-Chaplygin-Gas EoS \cite{Pourhassan:2014cea, Pourhassan:2014ika}
\begin{equation}
p = - \frac{B^2}{\rho} + A \rho,
\end{equation}
which is a good dark energy model unifying dust and cosmological constant, while at the same time, thanks to the barotropic term, the pressure can vanish at the surface of the star.

In this study we shall consider three different models studied before \cite{DEstars3,DEstars4,DEstars5}. The numerical values of the constant parameters $A,B$ are shown in Table I. Those values have been considered in previous works \cite{DEstars3,DEstars4,DEstars5}, as they ensure the existence of finite and well-behaved interior solutions capable of describing realistic astrophysical configurations with stellar masses and radii similar to those of neutron stars and strange quark stars. Although the numerical values of the parameters $A,B$ do not differ significantly from one model to another, they imply considerably different mass-to-radius relationships \cite{DEstars3,DEstars5}. 

\smallskip

Once the TOV equations have been integrated and the interior solution has been obtained, the moment of inertia, $I$, of non-rotating stars is computed by \cite{Panotopoulos:2018joc,Staykov:2014mwa}
\begin{equation}
I = \frac{8 \pi}{3} \: \int_0^R dr \: r^4 \: (p+\rho) \: \left( \frac{e^{\lambda/2}}{e^{\nu/2}} \right), \;  \; \; \; \; \; \bar{I}=\frac{I}{M^3},
\end{equation}
while $\bar{I}$ is the normalized moment of inertia. For rotating stars $J = I \: \Omega$, with $J$ being the angular momentum and $\Omega$ being the angular velocity, respectively, while the moment of inertia is computed by a slightly more complicated expression including $\Omega$, see for instance \cite{Murshid:2023xsw}. Next, we compute the function $Q(r)$ and we solve the Riccati differential equation to compute the tidal Love number, $k$, and the corresponding dimensionless deformability, $\Lambda$, in terms of $k$ and the factor of compactness $C=M/R$. 

\smallskip

After that we make a fit to obtain the function $Y(z)$, with $z \equiv ln(\Lambda)$ being the independent variable, and $Y \equiv ln(\bar{I})$ being the dependent variable. Given the large number of different EoSs that one may consider, it turns out that a single fit adequate for the range of quantities considered here is a fourth-order polynomial of the form
\begin{equation}
Y(z) = a + b z + c z^2 + d z^3 + e z^4,
\end{equation}
considered also in \cite{Yagi:2013bca,Yagi:2013awa,Yagi:2016bkt}, with the unknown coefficients $a,b,c,d,e$ to be determined. In Table II we show the values of the coefficients for the three models considered in this work. 

\smallskip

Finally, following \cite{inertia1,inertia2} we also fit the moment of inertia with the factor of compactness considering a fit of the form
\begin{equation}
I = M R^2 \: (a_2 + a_1 C + a_0 C^2),
\end{equation}
or equivalently of the form
\begin{equation}
\bar{I} = a_0 + \frac{a_1}{C} + \frac{a_2}{C^2},
\end{equation}
with the unknown coefficients $a_0,a_1,a_2$ to be determined. Given the large number of different EoSs that one may consider, it turns out that a single fit for the range of quantities considered here is the  

\smallskip

We find the following expressions:
\begin{equation}
\bar{I} = 5.543 + \frac{0.070}{C} + \frac{0.430}{C^2},
\end{equation}
for Model 1,
\begin{equation}
\bar{I} = 5.954 - \frac{0.039}{C} + \frac{0.437}{C^2},
\end{equation}
for Model 2, and
\begin{equation}
\bar{I} = 6.379 - \frac{0.157}{C} + \frac{0.444}{C^2},
\end{equation}
for Model 3.

\smallskip

All the numerical work has been performed with Wolfram Mathematica, and for each model the code runs, computes stellar properties and produces the plots within $\sim 55 sec$.

\smallskip

\begin{figure*}[ht!]
\centering
\includegraphics[scale=1.2]{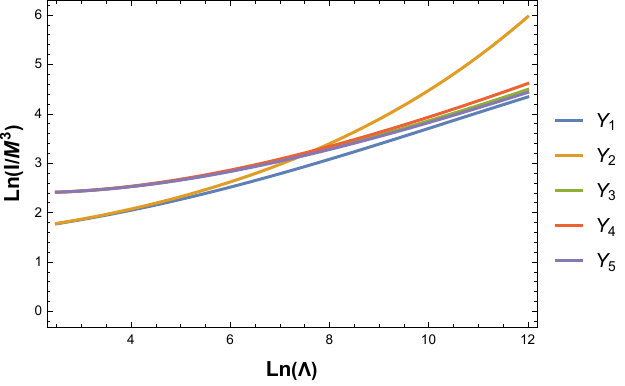} 
\caption{
Universal relations for the three dark energy stars studied here (functions $Y_3,Y_4,Y_5$), and universal relations found in previous works (function $Y_1$ from Table I of \cite{Yagi:2013awa} and function $Y_2$ from Table I of \cite{UR7}).
}
\label{fig:5} 	
\end{figure*}

In Figures \ref{fig:1}, \ref{fig:2} and \ref{fig:3} we have shown the normalized moment of inertia versus factor of compactness (lower panels) and versus dimensionless deformability (top panels) for all three models considered here. For comparison reasons, in Fig. \ref{fig:4} we have displayed the fitting functions for all three models considered in this work. It is observed that the curves are not distinguishable. To complete the triplet of relations, the lower panel of Fig. \ref{fig:4} shows the dimensionless deformability as a function of the factor of compactness. Similarly to the normalized moment of inertia, $\Lambda$, too, decreases with $M/R$. Since the other two functions $I(C)$ and $I(\lambda)$ are known, one basically has the function C-Love in parametric form, and so it is strightforward to obtain the function $C(\Lambda)$.
Finally, in Fig. \ref{fig:5} we have shown universal relations obtained in previous studies in order for the reader to be able to see how these relations compare between them and dark energy stars discussed in this work.

%%%%%%%%%%%%%%%%%%%%%%%%%%%%%TABLES%%%%%%%%%%%%%%%%%%%%%%%%%%%%%%%%

\begin{table}[ph!]
\centering
\caption{Values of the constant parameters $A,B$ for the three models considered in this work. 
%The numerical values of the parameters $M,q,b$ assumed to obtain the QNMs are the ones shown in Fig.~\ref{fig:1}.
}
{
\begin{tabular}{c|ccc} 
\toprule
Model  &  A  &  B 
\\ \colrule
\hline
1 & $\sqrt{0.4}$    &  $0.230 \times 0.001 \: km^{-2}$  \\ \hline
2 & $\sqrt{0.425}$  &  $0.215 \times 0.001 \: km^{-2}$  \\ \hline
3 & $\sqrt{0.45}$   &  $0.200 \times 0.001 \: km^{-2}$    \\ \hline
\botrule
\hline
\end{tabular} 
\label{table:First set}
}
\end{table}

%%%%%%%%%

\begin{table}[ph!]
\centering
\caption{Values of the coefficients of the fitting for the three models considered in this work. 
%The numerical values of the parameters $M,q,b$ assumed to obtain the QNMs are the ones shown in Fig.~\ref{fig:1}.
}
{
\begin{tabular}{c|ccccc} 
\toprule
Coefficient  &  a  &  b  &  c  &  d  &   e
\\ \colrule
\hline
Model 1 & 2.506 & -0.124 & 0.037 & -0.0013 & $1.877 \times 10^{-5}$   \\ \hline
Model 2 & 2.515 & -0.128 & 0.038 & -0.0013 & $1.965 \times 10^{-5}$   \\ \hline
Model 3 & 2.522 & -0.131 & 0.038 & -0.0014 & $2.061 \times 10^{-5}$  \\ \hline
\botrule
\hline
\end{tabular} 
\label{table:Second set}
}
\end{table}

%%%%%%%%%%%%%%%%%%%%%%%%%%%%END_TABLES%%%%%%%%%%%%%%%%%%%%%%%%%%%%%

%%%%%%%%%%%%%%%%%%%%%%%%%%%%%%%%%
\section{Discussion and summary}
%%%%%%%%%%%%%%%%%%%%%%%%%%%%%%%%%

To summarize our work, in the present article we have studied spherical configurations made of dark energy. The underlying equation-of-state was assumed to be the Extended-Chaplygin Gas equation-of-state characterized by two constant parameters, $A,B$, out of which $A$ is dimensionless and $B$ dimensionful. To be more precise, we have considered three different models, and the numerical values of the parameters were shown in Table I. First, we integrated the structure equations to obtain stellar interior solutions describing hydrostatic equilibrium, and we computed the factor of compactness, $C=M/R$, as well as the moment of inertia ($I$ and $\bar{I}=I/M^3$) of the non-rotating objects. Next, we solved the Riccati differential equation for the even metric perturbations to compute the gravito-electric tidal Love numbers, $k$, and the corresponding dimensionless deformabilities $\Lambda$. After that we made a fit to obtain the numerical values of the coefficients assuming a logarithmic function $\bar{I}(\Lambda)$. We also fitted the moment of inertia as a function of the factor of compactness. 

\smallskip 

\begin{figure*}[ht!]
\centering
\includegraphics[scale=0.8]{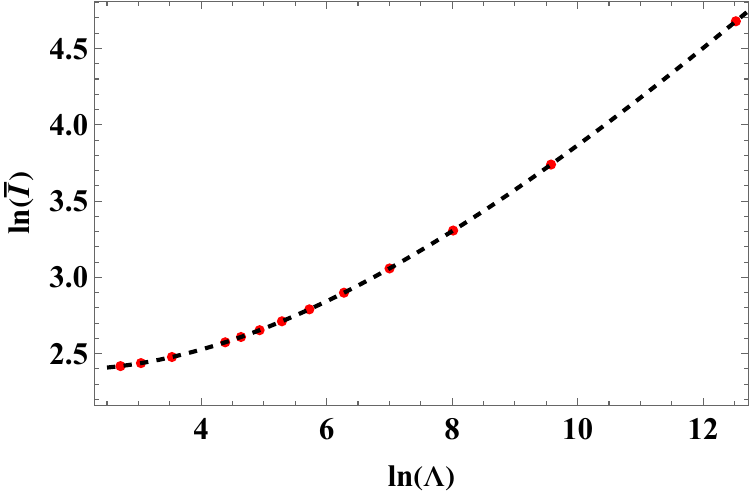} \ 
\includegraphics[scale=0.82]{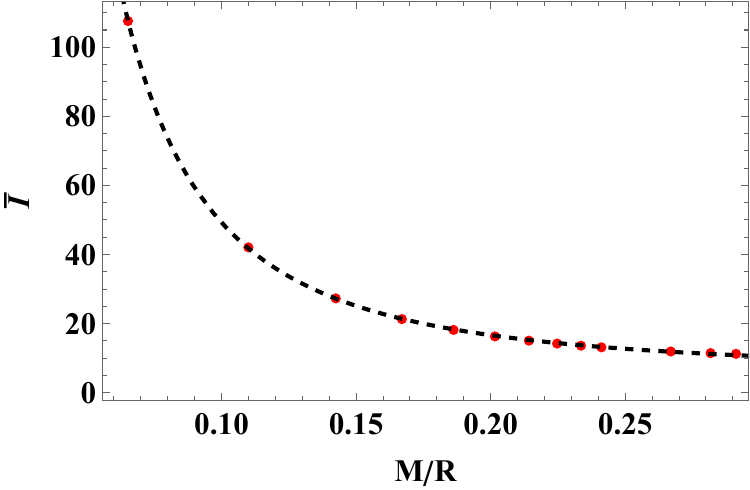}
\caption{
{\bf Upper panel:} Normalized moment of inertia, $\bar{I}=I/M^3$, versus dimensionless deformability for model 1. {\bf Lower panel:}  Normalized moment of inertia versus factor of compactness for model 1.
}
\label{fig:1} 	
\end{figure*}

\begin{figure*}[ht!]
\centering
\includegraphics[scale=0.8]{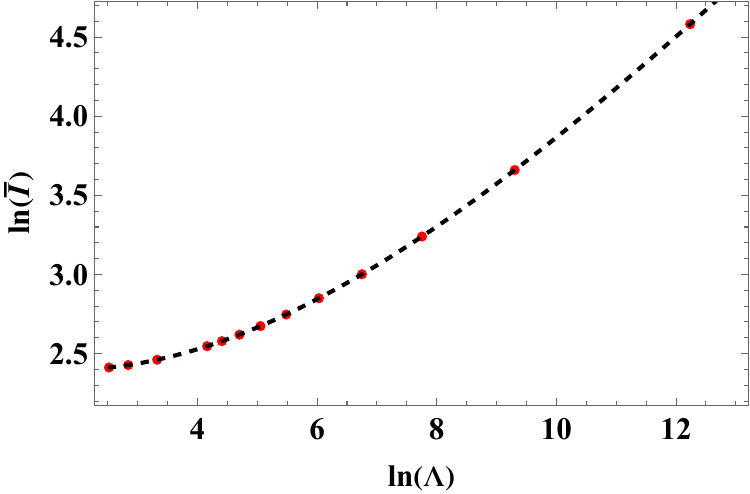} \ 
\includegraphics[scale=0.82]{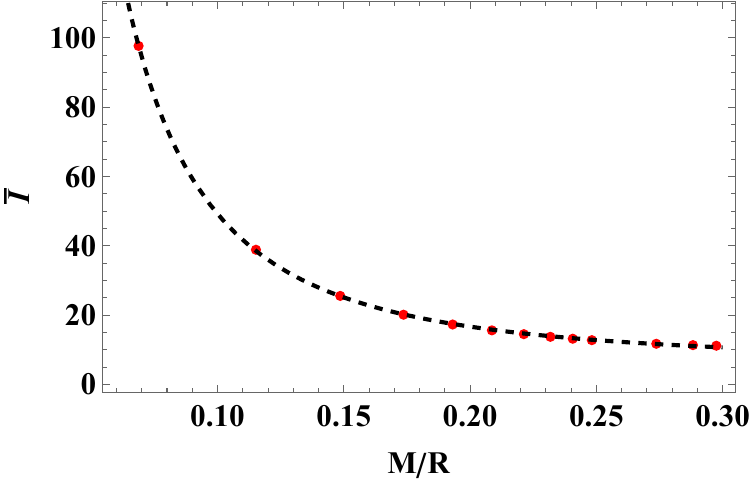}
\caption{
{\bf Upper panel:} Normalized moment of inertia versus dimensionless deformability for model 2. {\bf Lower panel:}  Normalized moment of inertia versus factor of compactness for model 2.
}
\label{fig:2} 	
\end{figure*}

\begin{figure*}[ht!]
\centering
\includegraphics[scale=0.8]{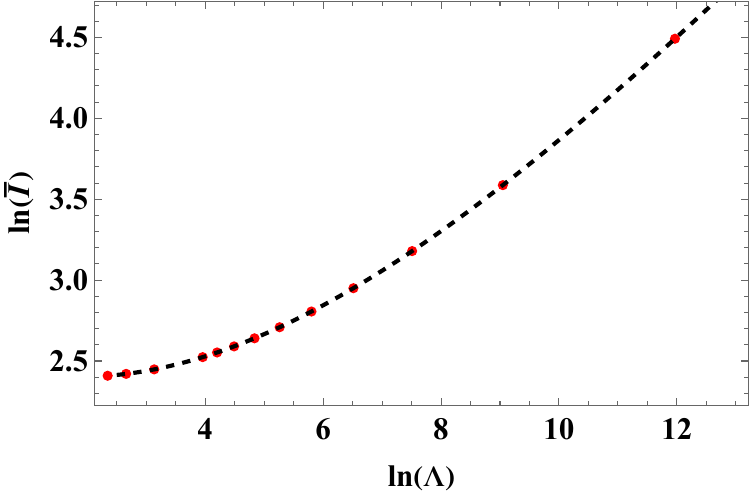} \ 
\includegraphics[scale=0.82]{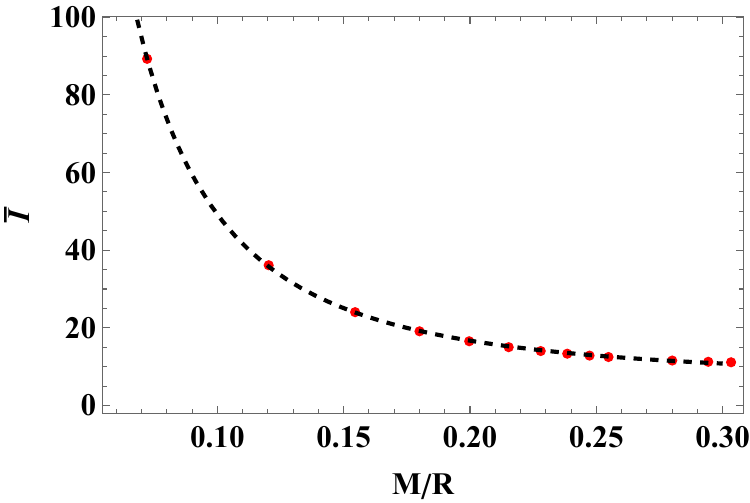}
\caption{
{\bf Upper panel:} Normalized moment of inertia versus dimensionless deformability for model 3. {\bf Lower panel:}  Normalized moment of inertia versus factor of compactness for model 3.
}
\label{fig:3} 	
\end{figure*}

\begin{figure*}[ht!]
\centering
\includegraphics[scale=0.83]{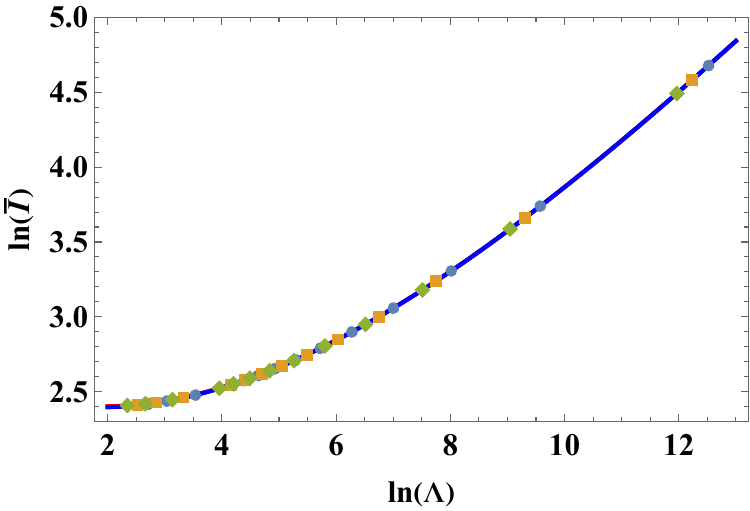} \ 
\includegraphics[scale=0.88]{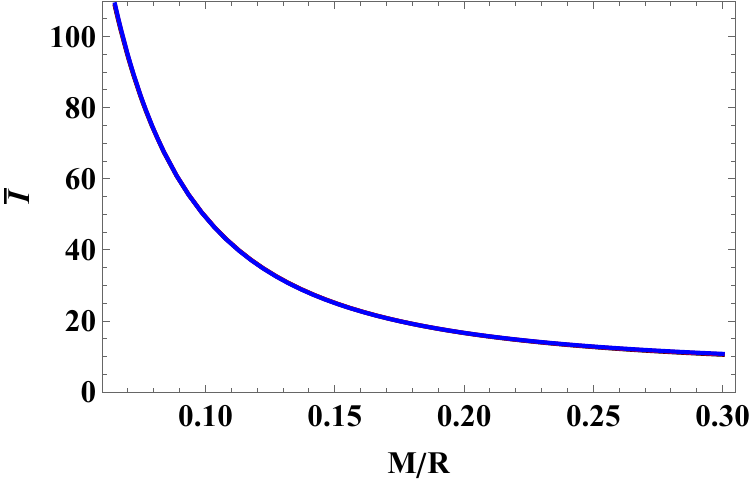} \
\includegraphics[scale=0.88]{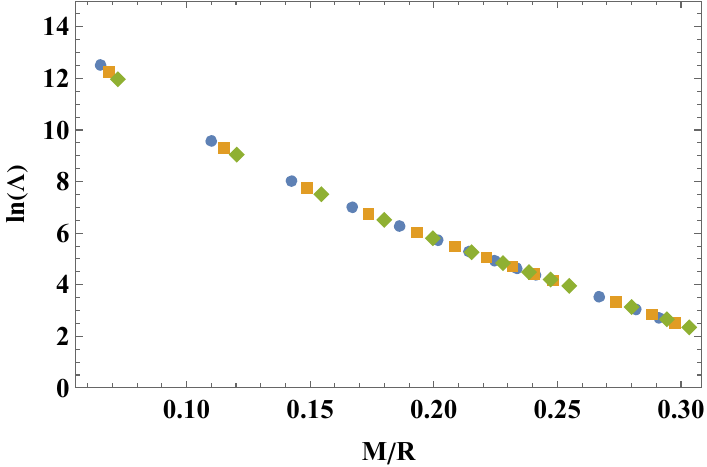}
\caption{
Triplet of relations discussed in this work. {\bf Upper panel:} Normalized moment of inertia versus dimensionless deformability for all three models (Model 1: Blue circles, model 2: orange squares, model 3: green rhombus) considered here, both numerical points and curves of the fitting functions. The three curves (model 1 in black, model 2 in red, model 3 in blue) are not distinguishable. {\bf Middle panel:} Normalized moment of inertia versus factor of compactness for all three models considered in this work (model 1 in black, model 2 in red, model 3 in blue). Similarly to the other panel, the three curves of the fitting functions are not distinguishable. {\bf Lower panel:} Dimensionless deformability versus factor of compactness for all three models (Model 1: Blue circles, model 2: orange squares, model 3: green rhombus).
}
\label{fig:4} 	
\end{figure*}

Our numerical results were summarized in Table II and in Figures \ref{fig:1}, \ref{fig:2}, \ref{fig:3}, and \ref{fig:4}. For all three models considered here, $\bar{I}$ increases with the dimensionless deformability (top panels), and decreases with $C$ (lower panels). Regarding the $I-C$ relations, all coefficients were found to be positive in the case of model 1, whereas in the case of models 2 and 3 the coefficient $a_1$ was found to be negative and the others positive. Regarding the $I-\Lambda$ relations, for all three models considered here the coefficient $b,d$ were found to be negative, and the other three coefficients were found to be positive. This is to be contrasted to the results reported in \cite{Yagi:2013bca,Yagi:2013awa}, where only the coefficient $e$ was found to be negative. Moreover, it is observed that each coefficient is one order of magnitude lower than the previous one. Finally, for comparison reasons, in Fig. \ref{fig:4} we have displayed the fitting functions for all three models considered in this work. It is observed that the curves are not distinguishable, and therefore our findings indicate that the form of the functions does not depend on the details of the underlying equation-of-state.

%%%%%%%%%%%%%%%%%%%%%%%%PLOTS%%%%%%%%%%%%%%%%%%%%%%%%%%%%%%%%%

%%%%%

%%%%%

%%%%%%

%%%%%

%%%%%%%%%%%%%%%%%%%END_FIGURES%%%%%%%%%%%%%%%%%%%%%%%%%%%%%%%

\section{Acknowledgments}

The author wishes to thank the anonymous referees for useful comments and suggestions, and N. Dimakis for helping with the figure \ref{fig:5}.

\end{document}